\documentclass[11pt]{article}

\newcommand{\ed}{\ \stackrel{d}{=} \ }

\newcommand{\FF}{\mbox{${\cal F}$}}

\newcommand{\eps}{\varepsilon}

\newcommand{\bE}{\mathbf{E}}

\newcommand{\bT}{\mathbf{T}}

\newcommand{\bV}{\mathbf{V}}

\newcommand{\var}{{\rm var}\ }

\newcommand{\sfrac}[2]{{\textstyle\frac{#1}{#2}}}

\usepackage{amssymb}
\usepackage{epsfig}
\newcommand{\flow}{\mathbf{f}}

\newcommand{\Zbold}{{\mathbb{Z}}}
\newcommand{\cfpp}{c_{\mbox{{\tiny FPP}}}}
\renewcommand{\root}{\mbox{\tiny{root}}}

\newcommand{\CD}{customer driven\ }
\newcommand{\vw}{w}
\author{David J. Aldous\thanks{Research supported by N.S.F. Grant
DMS0203062}
\\
\\
\\
       University of California\\
       Department of Statistics\\
        367 Evans Hall \# 3860\\
       Berkeley CA 94720-3860}

\title{Cost-Volume Relationship for Flows Through a Disordered Network}
\begin{document}
\maketitle

\begin{abstract}
In a network where the cost of flow across an edge is nonlinear in the
volume of flow,
and where sources and destinations are uniform,
one can consider the relationship between
total volume $v$ of flow through the network and the minimum cost $c = \Psi(v)$
of any flow with volume $v$.
Under a simple probability model
(locally tree-like directed network, independent cost-volume functions for different
edges)
we show how to compute $\Psi(v)$ in the infinite-size limit.
The argument uses a probabilistic reformulation of the cavity method
from statistical physics,
and is not rigorous as presented here. 
The methodology seems potentially useful for many problems concerning
flows on this class of random networks. 
\end{abstract}

{\em Key words:} 
cavity method,
network flow,
probability model.

{\em MSC2000 subject classification:}  
Primary: 90B15

{\em OR/MS subject classification:} 
Primary: Networks/graphs

\newpage
\section{Introduction}
The time (``cost") it takes you to drive a given segment of
road depends on the amount (``volume") of traffic,
increasing as volume increases up to some critical value at which
the road becomes jammed.
So there's a cost-volume curve for each road segment.
Now consider the road network of a city, 
with many vehicles simultaneously travelling
from different ``sources" to different destinations,
using minimum-cost routes depending on congestion pattern.
As we linearly scale the overall volume of traffic,
the average cost-per-vehicle will also increase
as volume increases, up to some critical value at which
the network becomes jammed. 
So there's a cost-volume curve for the network as a whole 
(depending also on the source-destination pattern).

This paper gives a foundational mathematical study of the idea above,
in an artificially simple setting. 
One can view this topic as akin to statistical physics: we seek to
understand how the ``macroscopic" behavior of the network
(the network cost-volume function) emerges from the ``microscopic"
specification (a probability distribution on edge cost-volume functions
and a probability distribution on network topology).  
And our methodology is a recent probabilistic reformulation of the
{\em cavity method} of statistical physics.  
Apparently this is the first paper to apply such methodology to
explicitly ``network flow" problems, and it is plausible that a
broader range of problems than treated here could be studied by the
same methodology, albeit with some intrinsic caveats
noted in section \ref{sec-purpose}.

Of course, the study of flows in networks is a centerpiece of classical 
Operations Research and has evident applications in several Engineering
disciplines
\cite{AMO93}.
But we don't know any work which is closely related to the present paper,
so we will defer literature discussion until a later survey paper 
intended to present a much broader view of the topic of flows through random networks.
We should emphasize that we are discussing deterministic flows
on random networks, in contrast to 
{\em queueing theory} which studies random
flows on deterministic networks: see \cite{kelly-NR} for a brief survey of routing questions
within that setting.

\subsection{A network model}
\label{sec-anm}
{\bf The random layer graph model.}
Take $M$ layers, each with $N$ vertices.
For each $1 \leq i \leq M-1$ create directed edges from some vertices in
layer $i$ to some vertices in layer $i+1$.
The choice of edges is uniform random, subject to the constraint
\begin{quote}
each layer-$i$ vertex has out-degree $2$, and
each layer-$(i+1)$ vertex has in-degree $2$.
\end{quote}
This defines a random graph with
$ MN$ vertices $w$ and with $2(M-1)N$ directed edges $e$.  
See Figure 1.

\setlength{\unitlength}{0.5in}
\begin{picture}(5,4)(-1.5,-0.5)
\multiput(0,0)(1,0){6}{\circle*{0.2}}
\multiput(0,1)(1,0){6}{\circle*{0.2}}
\multiput(0,2)(1,0){6}{\circle*{0.2}}
\multiput(0,3)(1,0){6}{\circle*{0.2}}
\put(0,0){\line(2,1){2}}
\put(0,0){\line(4,1){4}}
\put(1,0){\line(-1,1){1}}
\put(1,0){\line(2,1){2}}
\put(2,0){\line(-2,1){2}}
\put(2,0){\line(3,1){3}}
\put(3,0){\line(0,1){1}}
\put(3,0){\line(2,1){2}}
\put(4,0){\line(0,1){1}}
\put(4,0){\line(-3,1){3}}
\put(5,0){\line(-4,1){4}}
\put(5,0){\line(-3,1){3}}
\put(0,1){\line(0,1){1}}
\put(0,1){\line(3,1){3}}
\put(1,1){\line(1,1){1}}
\put(1,1){\line(3,1){3}}
\put(2,1){\line(-1,1){1}}
\put(2,1){\line(3,1){3}}
\put(3,1){\line(-3,1){3}}
\put(3,1){\line(-1,1){1}}
\put(4,1){\line(-3,1){3}}
\put(4,1){\line(1,1){1}}
\put(5,1){\line(-2,1){2}}
\put(5,1){\line(-1,1){1}}
\put(0,2){\line(1,1){1}}
\put(0,2){\line(5,1){5}}
\put(1,2){\line(1,1){1}}
\put(1,2){\line(2,1){2}}
\put(2,2){\line(-2,1){2}}
\put(2,2){\line(2,1){2}}
\put(3,2){\line(-2,1){2}}
\put(3,2){\line(0,1){1}}
\put(4,2){\line(-4,1){4}}
\put(4,2){\line(0,1){1}}
\put(5,2){\line(-3,1){3}}
\put(5,2){\line(0,1){1}}
\end{picture}

{\bf Figure 1.}
{\small 
A realization of the random layer graph with $M = 4, \ N = 6$.
}

\vspace{0.09in}
\noindent
For our purposes the key feature of this model
is that as $M,N \to \infty$ the sequence of random layer graphs
satisfies
{\em local weak convergence}
to the infinite tree
$\bT$ in which each vertex has in-degree $2$ and out-degree $2$.
Local weak convergence \cite{me101} means:
\begin{quote}
Take a uniform random vertex to be a root of the $n$-vertex graph.
As $n \to \infty$, the subgraphs on vertices within an arbitrary fixed
distance (number of edges) from the root converge in distribution
to the corresponding subgraph of the limit $\bT$,
considered as a rooted graph.
\end{quote}
Now suppose that on each edge $e$ of the random layer graph
there is a 
function $(\Phi(e,v), \ v \geq 0)$ representing the cost of a flow of
volume $v$ across $e$.  Equivalently, consider the cost-per-unit-flow
$\phi(e,v) = v^{-1}\Phi(e,v)$.
(Note: the mathematics works more cleanly with ``total cost" functions
like $\Phi$ and $\Psi$ below, but the interpretation is more
intuitive in terms of cost-per-unit-volume functions $\phi$ and $\psi$.)
Suppose we wish to send flow of volume $v_{M,N}$ 
through the network, i.e. from layer $1$ to layer $M$, along directed edges.
Each possible such ``global flow"
has some total cost, and so one can seek to study the minimum total cost
as a function of volume 
$v_{M,N}$,
under some model of edge-costs.

\vspace{0.09in}
\noindent
{\bf The edge-cost model.}
Fix a probability distribution on functions $\Phi(v)$
(equivalently: on functions $\phi(v) = v^{-1}\Phi(v)$).
For each edge $e$ of the random layer graph,
let $\Phi(e,v)$ be chosen independently from this probability distribution.

Discussing $M,N \to \infty$ limits involves scaling conventions,  
whose details
we specify here but which (as described below) 
are easily interpretable without these details.
Because there are $2N$ edges 
between successive layers,
the typical flow per edge will be order 
$v_{M,N}/(2N)$.
We therefore take ``standardized volume" 
$0<v<\infty$ and set $v_{M,N} = 2Nv$.
With the resulting order $1$ flows through edges, the total
cost will scale as the number $2N(M-1)$ of edges.
Thus we define
{\em standardized cost of the optimal flow with standardized volume $v$}
to be 
\[
\Psi_{M,N}(v) = \sfrac{1}{2N(M-1)} \ 
\mbox{({\small minimal cost over flows of volume $2Mv$ 
through the network})}
. \]
The function $\Psi_{M,N}(v)$ is random because it depends on the realizations
of the graph and 
of edge-flow functions, but by virtue of the standardization
we expect a deterministic limit function $\Psi$:
\[
\Psi_{M,N}(v) \to \Psi(v) 
\mbox{ in probability, } 
\quad 0<v<\infty  
\]
as $M,N \to \infty$ with not too dissimilar orders of magnitude.
Set $\psi(v) = v^{-1}\Psi(v)$.  
To interpret the limit function more intuitively, 
consider as a benchmark the ``uniform" flow of constant volume $v$
along each edge.
This has normalized volume $v$ and limit normalized cost
$E \phi(v)$.
The purpose of the standardizations is simply to be able to compare cost of the optimal
flow of given volume in our model with the  
cost of the uniform flow of the same volume.

The setting where edges have some finite {\em capacity} 
(maximum allowed volume) fits our setup by taking
$\Phi(v) = \infty$ for $v$ larger than the capacity.
In this case we expect the network has some finite maximum
standardized volume $v^*$:
\begin{eqnarray}
\Psi(v) & < \infty,& \ v < v^* \nonumber\\
&=\infty,& \ v > v^* . \label{qqq}
\end{eqnarray}
Note that $v^*$ will not depend on edge-costs, just on edge-capacities.

\subsection{Methodology}
\label{sec-purpose}
The purpose of this paper is to point out that it is indeed possible to analyze
the model above.
That is, one can via theoretical arguments obtain the limit network
cost-volume function $\Psi(v)$ 
for a given distribution of edge cost-volume functions.
The results are presented in section \ref{sec-results} in a variety of particular cases
and specializations.

To be upfront about the caveats:
\begin{itemize}
\item The arguments in this paper are non-rigorous.
\item The methodology deals with limits as number of vertices grows to infinity, and is only useful when the underlying graphs are
``locally tree-like" (local weak convergence to some infinite tree, maybe random).
\item Getting explicit results involves numerical solution of a 
fixed-point equation (RDE) for an unknown probability distribution.
\end{itemize}
{\small 
While the latter two caveats are intrinsic to the methodology, 
the first is more a technical matter: we 
understand conceptually what steps
are needed to make a rigorous proof,
but implementing details of some of the steps in general settings seems
astronomically far out of reach of current theory.
A high-level description of the methodology, which
we describe as 
``probabilistic reformulation of the cavity method", 
can be found in \cite{me107} section 7.5.
In this paper our focus is on exhibiting the calculations (section \ref{sec-methods}) and their
results in the network flow setting, without attempting
rigorous justification.  However, one of our simpler examples
(maximum density of edge-disjoint
infinite paths in a randomly obstructed 
infinite tree; section \ref{sec-obstruct})
provides an appealing benchmark problem for future development of
rigorous proofs.

We postpone further discussion until sections \ref{sec-formalize} and \ref{sec-discuss}.
}

\section{Results}
\label{sec-results}
Within the model of section \ref{sec-anm},
we will describe the network cost-per-unit-volume curve $c = \psi(v)$ in five examples
at varying levels of generality.  
How these results are derived will be explained in section \ref{sec-methods}.

\subsection{A traffic flow model}
\label{sec-traffic}
The real-world relationship between traffic speed and traffic
density has of course been studied in detail; see \cite{gazis} for
an introduction to this theory.  Let us take the most naive model
in which speed $s$ is a decreasing linear function of traffic density $\rho$:
\[ s = s_0(1 - \alpha \rho ). \]
Note that our flow volume $v$ equals $s \rho$.
This model implies there is a maximum possible flow volume, attained
at speed $s_0/2$.
In our setting, ``cost" $c$ is traversal time, that is proportional
to $1/s$.  
Solving for $c$ in terms of $v$ gives
the cost-volume function 
for an edge:
\begin{eqnarray}
c = \phi(v) &=& c_0 \ 
\frac{1 - (1 - \frac{v}{\vw_*})^{1/2}}{\frac{v}{2\vw_*}} \quad
 ; v \leq \vw_*
\label{cv-para}\\
&=& \infty, \quad v > \vw_*
\nonumber
\end{eqnarray}
where $c_ 0 = \phi(0+)$ is the cost-per-unit-volume at the zero volume limit,
and $\vw_*$ is the maximum volume.
Also, the cost-per-unit-volume at maximum flow equals $2c_0$.

To make a probability model we take $c_0 = 1$ and let 
$\vw_*(e)$ be independent over edges $e$
with 
Exponential($1$) distribution.
Figure 2 shows the network cost-per-unit-volume curve
$c = \psi(v)$.

\psfig{figure=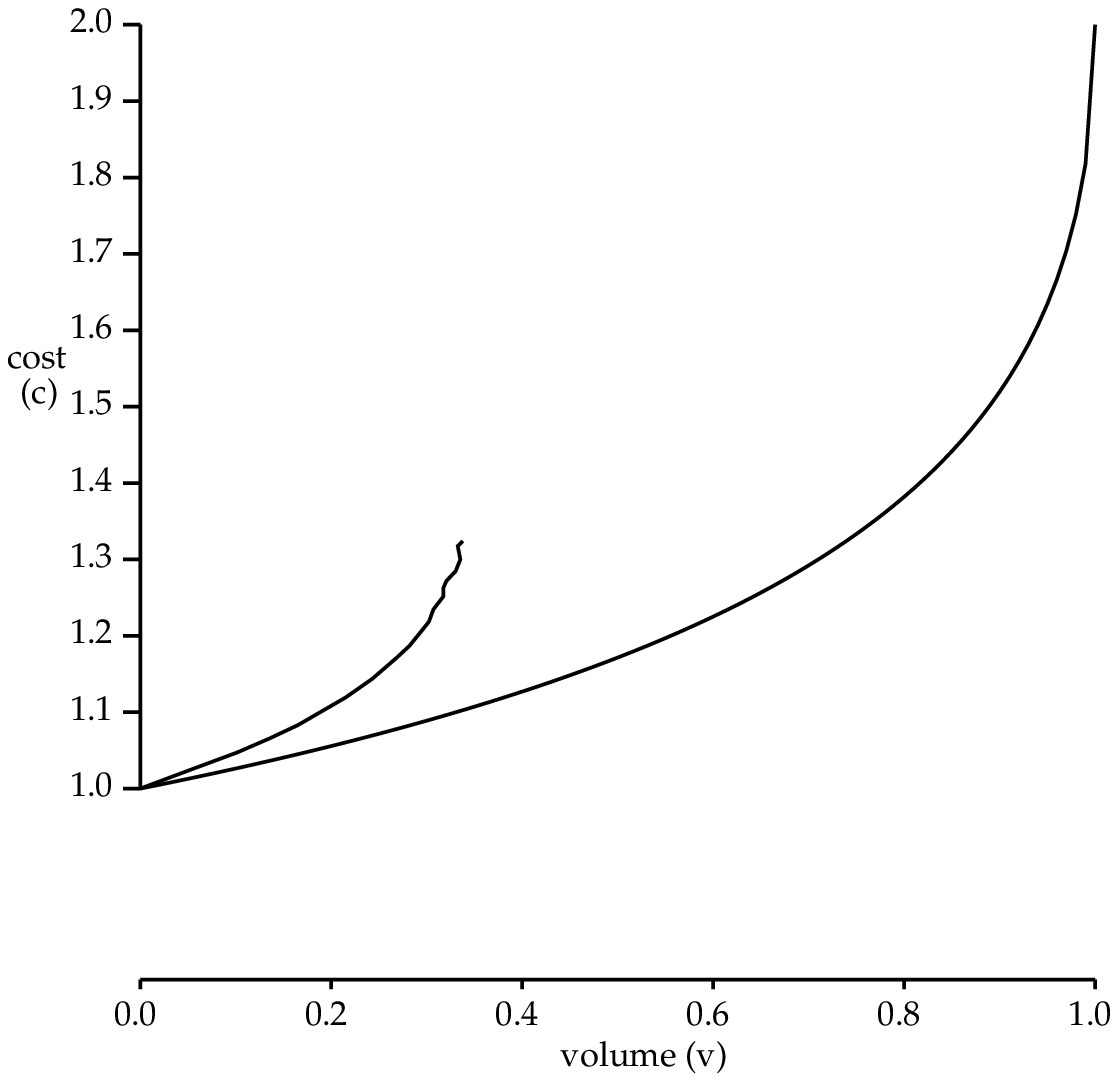}

{\bf Figure 2.}
{\small 
The long curve is the edge cost-per-unit-volume function $c = \phi(v)$ at (\ref{cv-para})
with $c_0 = \vw_* = 1$.
The short curve is the network cost-per-unit-volume function $c = \psi(v)$.
Numerical results from bootstrap Monte Carlo solution of the 
RDE (\ref{Xtrde}).
Irregularities are artifacts of sampling variation, as explained in section 
\ref{sec-bootstrap}.
}

\vspace{0.09in}
\noindent
Because each edge $e$ has $\phi(e,0) = 1$ 
we obviously have $\psi(0+) = 1$.
The maximum normalized volume of network flow is numerically
about $0.34$ and the corresponding cost-per-unit-volume
is numerically about $1.33$.
The network cost-volume curve has the same qualitative shape as
the edge cost-volume curve.

\subsection{Capacity constraints}
\label{sec-capac}
As mentioned before, the case where edges $e$ have a
maximum capacity $K(e)$ can be fitted into our framework by
assigning infinite cost to larger flows.

\psfig{figure=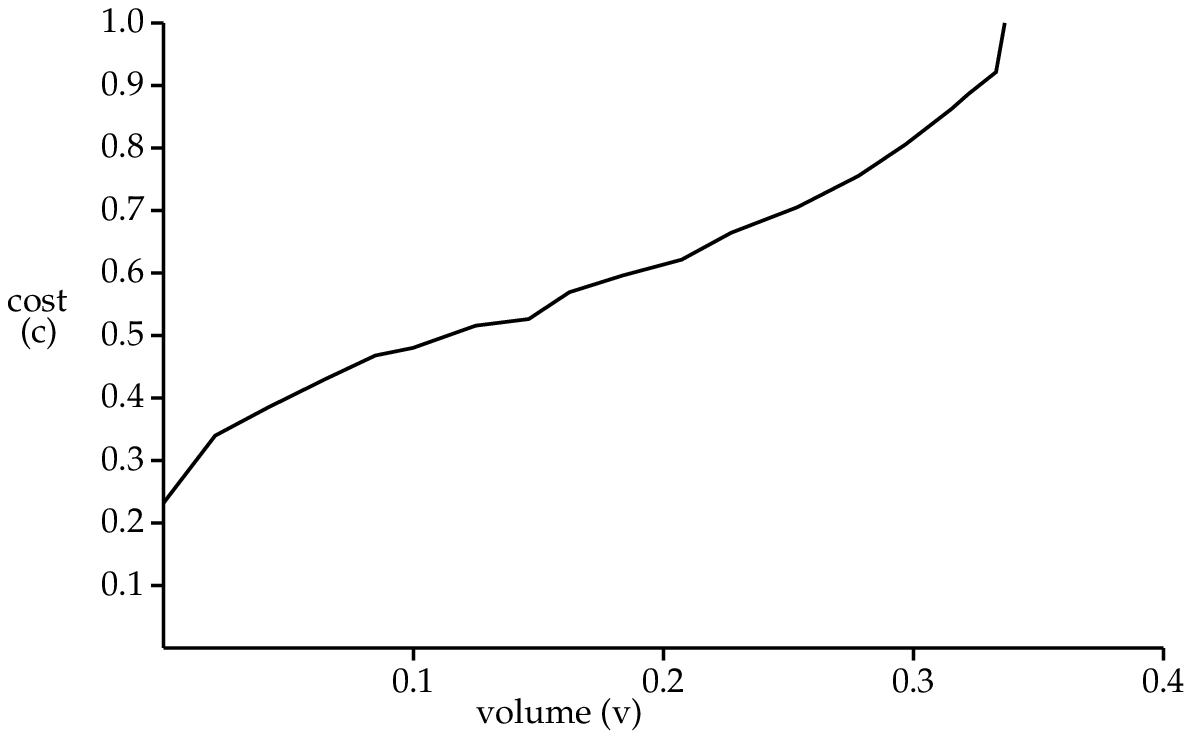}

{\bf Figure 3.}
{\small 
The network cost-per-unit-flow
function $c = \psi(v)$, in the case where 
the edge cost-per-unit-flow
is a constant $C(e)$ up to a capacity $K(e)$, where
$(C(e),K(e))$ are independent Exponential ($1$) as $e$ varies.
Numerical results from bootstrap Monte Carlo solution of the 
RDE (\ref{Xtrde}). 
Irregularities are artifacts of sampling variation, as explained in section 
\ref{sec-bootstrap}.
}

\vspace{0.09in}
\noindent
Taking cost-per-unit-flow to be constant up to the capacity
gives 
\begin{eqnarray*}
\phi(e,v) &=& C(e) \quad 0 \leq v \leq K(e)   \\
&=& \infty \quad v > K(e)  \nonumber
\end{eqnarray*}
where $(C(e),K(e))$ are i.i.d. as $e$ varies.
We treat the example where $C(e)$ and $K(e)$
are independent with Exponential($1$) distribution.
Figure 3 shows the network
cost-per-unit volume function
$\psi(v)$.

Some aspects of this curve are understandable by theory.
Specializing a general large deviation result (\ref{eq-brw}) to the 
Exponential($1$) case shows that 
\begin{equation}
\mbox{
 $\psi(0+) \approx 0.23196$
is the solution of 
$x - \log x = 1 + \log 2$} . \end{equation} 
The maximal volume $v^*$ must be the same as in the previous
example (numerically, about $0.34$).  Because the 
cost-per-unit-flow on an edge is independent of edge capacity and has mean $1$,
we must have 
$\psi(v^*) = 1$.

\subsection{Unit edge capacities and the scaling exponent
in mean-field first passage percolation}
\label{sec-scaling}
This is the first of two specializations in which we take the 
edge-capacities to be constant 
($K = 1$);
section \ref{sec-UEC} explains how this leads to some mathematical simplification.
In this section, we specialize the model of the previous section 
to the case $K(e) = 1$
of constant edge capacities.
That is,
\begin{eqnarray}
\phi(e,v) &=& C(e) \quad 0 \leq v \leq 1   \label{unit-exp}\\
&=& \infty \quad v > 1  \nonumber
\end{eqnarray}
where the $C(e)$ are i.i.d. with Exponential($1$) distribution.
As in the previous section, we know from (\ref{eq-brw})
that the low-volume limit of the network cost function is
\[c = \psi(v) \downarrow \psi(0+) \approx 0.23196 \mbox{ as } v \downarrow 0 . \]
Here we examine in more detail the cost-volume curve in the low-flow regime.
Rewrite $\psi(0+)$ as $\cfpp$, 
to emphasize its interpretation as the time constant for
first passage percolation 
(see section \ref{sec-linear}).
To make an analogy below with percolation functions, we
consider the inverse function
$v = \psi^{-1}(c)$
giving volume as a function of cost-per-unit-volume.
Table 1 gives numerical results 
in the low-flow regime.

\vspace{0.2in}

$\begin{array}{cccccccc}
\lambda&&0.280&0.300&0.320&0.340&0.360&0.380\\
\mbox{cost } c&&0.267&0.279&0.290&0.302&0.313&0.327\\
\mbox{volume } v&&0.013&0.027&0.046&0.067&0.086&0.109\\
12.7(c - \cfpp)^2&&0.015&0.028&0.043&0.063&0.084&0.115
\end{array}$

\vspace{0.1in}

{\bf Table 1.}
{\small 
Volume and cost-per-unit-volume relationship for model
(\ref{unit-exp}) in the low volume regime.
Numerical results from bootstrap Monte Carlo solution of the 
RDE (\ref{Xrde}), showing a good fit
to $v = \psi^{-1}(c) \sim 12.7 (c - \cfpp)^2$.
The $\lambda$ is a parameter used to construct $c$ as an implicit function
of $v$.
}

\vspace{0.09in}
\noindent
Recall \cite{gri99} that in classical site or bond percolation
on $\Zbold^d$ there is a {\em percolation function}
\[ f(p) = P(\mbox{ origin in some infinite component}) \]
where $p$ is the underlying probability on sites or bonds.
There is a critical point $p_*$ at which $f(\cdot)$ becomes non-zero,
and considerable attention has been paid to 
{\em scaling exponents} 
\[ f(p) \asymp (p - p_*)^\alpha  \mbox{ as } p \downarrow p_* .\]
We do not know any parallel discussion in the setting of first passage
percolation, but our setup suggests one possible formulation.
In the model above, the inverse function
$v = \psi^{-1}(c)$ of $c = \psi(v)$
satisfies (Table 1)
\begin{equation}
 \psi^{-1}(c) \asymp (c - 
\cfpp )^2 . \label{scale-fpp}
\end{equation}
Our model can be viewed as the ``mean-field'' analog of oriented
first passage bond percolation on $\Zbold^d$.
In the latter model, giving edges capacity $1$ and interpreting the random
edge-traversal times as costs, we can define a cost-volume curve
as in this paper, and presumably one gets dimension-dependent 
scaling exponents in (\ref{scale-fpp}).
This seems an interesting, though difficult, topic for future research.

\subsection{Maximal flow through the randomly obstructed networks}
\label{sec-obstruct}
Fix $1/2<p<1$.
Consider the case where edge-costs are constant
($C = 1$)
and where the edge-capacities are either $0$ or $1$:
\[ P(K = 1) = p; \quad P(K =0) = 1-p . \]
In other words, a proportion $1-p$ of edges are obstructed
and permit zero flow.
In this model, the cost-volume curve is not an issue, 
since
normalized cost per unit volume is just $1$.
However, it is natural to ask how the maximum normalized volume $v^* = v^*(p)$ at (\ref{vstar})
behaves as a function of $p$.

Note that we may reformulate the model by taking $K = 1$
(all edges present with unit capacity) and taking
\begin{equation}
 P(C = 1) = p; \quad P(C = \infty) = 1-p  \label{Cflow}
\end{equation}
which has the same effect of eliminating from consideration
a proportion $1-p$ of edges.
As mentioned before, the case of constant edge-capacity 
is mathematically simpler.

The curve $v^*(p)$ is shown in Figure 4.
The qualitative endpoint behavior observed numerically
is not hard to understand theoretically -- see section \ref{sec-ROends}.

\psfig{figure=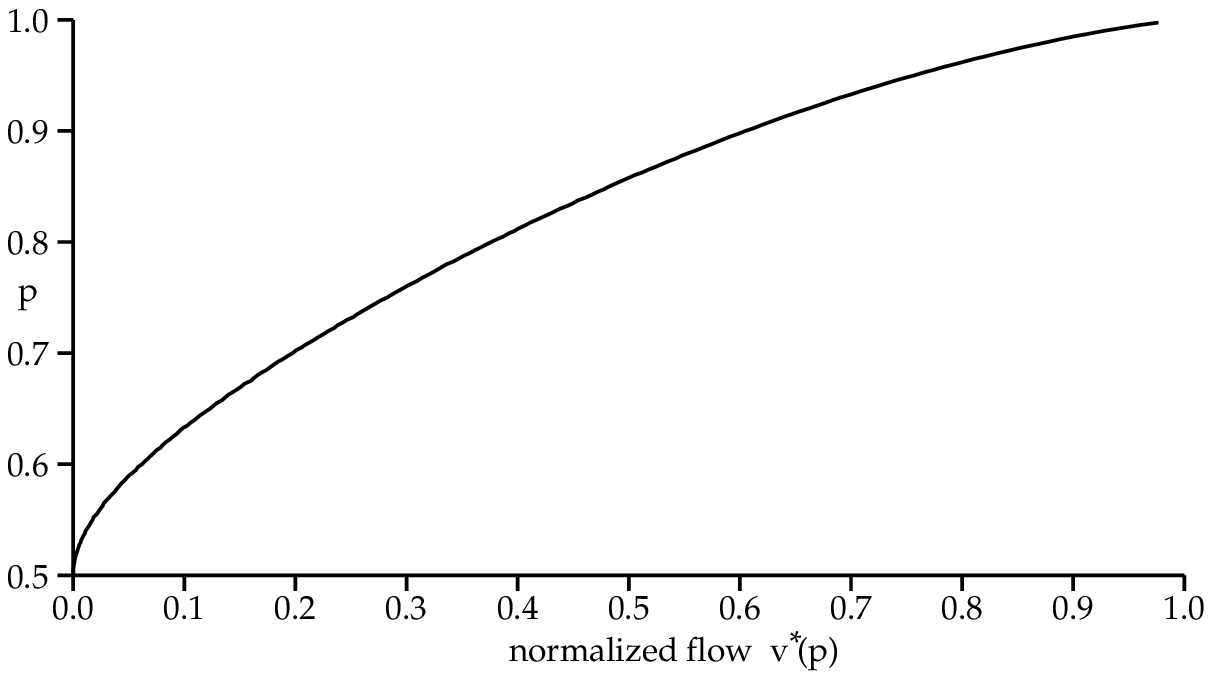}

{\bf Figure 4.}
{\small The relationship between $p$ and $v^*(p)$
in the randomly obstructed network model.
The curve was obtained by bootstrap Monte Carlo solution of
the RDE 
(\ref{Zrde}).
The endpoint behavior is
$v^*(p) \approx 0.76(p - 1/2)^2$ as $p \downarrow 1/2$;
$1 - v^*(p) \approx 1.56 (1-p) \log(\frac{1}{1-p})$ as $p \uparrow 1$.
}

\subsection{Quadratic costs}
\label{sec-quadratic}
Consider the case
\[ \Phi(v,e) = \kappa(e) v^2 \quad 
\mbox{ that is }
\phi(v,e) =  \kappa(e) v  \]
where $\kappa(e)$ is i.i.d. over edges $e$.
This has the obvious scaling property that if
$\flow$ has 
$v(\flow) = v_0, \ c(\flow) = c_0$
then a scaled flow $\alpha \flow$ has
$v(\alpha \flow) = \alpha v_0, \ c(\alpha \flow) = \alpha^2 c_0$.
So the network cost-per-unit-volume function must be of the form
\[
c= \psi(v) = \bar{\kappa} v
\]
where $\bar{\kappa}$
depends on the distribution of $\kappa(e)$.
Our normalization convention ensures
\[ \mbox{ if $P(\kappa(e) = 1) = 1$ then $\bar{\kappa} = 1$ . } \]

Figure 5 shows numerical results in the case where
$\kappa(e)$ has  
Gamma($a,a$) distribution 
(recall this has mean $a$ and standard deviation $a^{-1/2}$).
In the $a \to \infty$ limit we have $\kappa(e) = 1$
and so $\bar{\kappa} = 1$.
The ``myopic" flow 
with volume $1$ across each edge
always has normalized cost $1$.
As $a$ decreases, the variability
of $\kappa(e)$ increases and this causes the
normalized cost $\bar{\kappa}$ of the optimal flow to decrease,
because flow can take advantage of cheaper edges.

\psfig{figure=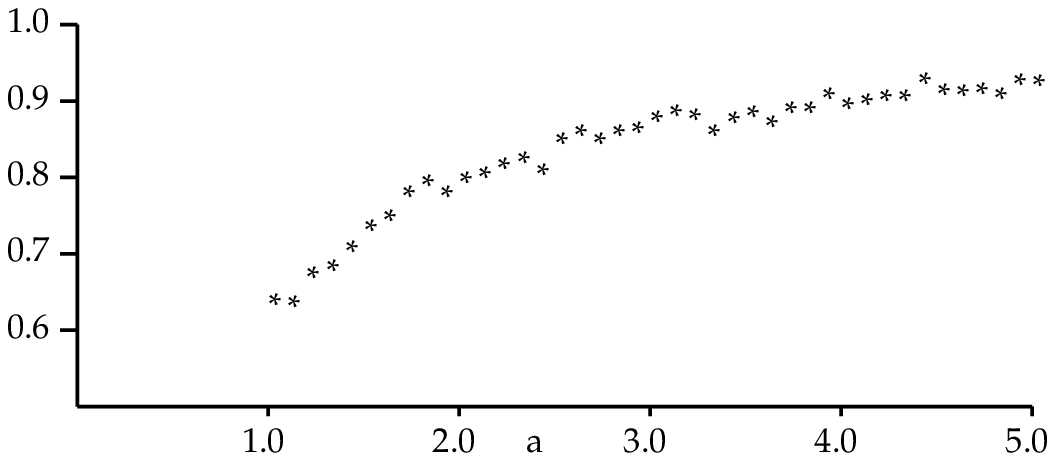}

{\bf Figure 5.}
{\small 
The case of quadratic 
costs with Gamma($a,a$) distribution, 
$1 \leq a \leq 5$.
The curve shows $\bar{\kappa}(a)$ as a function of $a$.
Numerical results from bootstrap Monte Carlo solution of the 
RDE (\ref{Xtrde}).
Irregularities are artifacts of sampling variation, as explained in section 
\ref{sec-bootstrap}.
}

\vspace{0.09in}
\noindent
Implementing the optimal flow through the network would require centralized routing 
policy.
It is natural to compare this to decentralized routing schemes,
and (viewing the network as traffic flow on an infinite tree) the natural
decentralized scheme is to have customers leaving each vertex
choose the cheaper out-edge to traverse next 
(this of course depends on the flows of other customers).
This {\em \CD} scheme turns out to be comparatively simple to analyze
(in the infinite tree limit).
Note that, compared to the myopic scheme, 
the \CD scheme benefits from being able to put more flow 
through the cheaper out-edge; on the other hand, the fact
that the volume of flow through different vertices is non-uniform
will tend (in the ``quadratic" setting) to increase costs.
Working through the analysis (section \ref{sec-greedy})
gives the remarkable conclusion that in the present setting
(quadratic costs; Gamma distribution of $\kappa(e)$) 
the normalized cost of the \CD scheme is exactly $1$, the same
as the myopic scheme.
We have no non-calculational explanation of this intriguing result.
Also, in this setting the applicability of infinite-tree analysis
to finite network problems is somewhat problematical.

\section{Implementing the cavity method}
\label{sec-methods}
\subsection{The infinite network model}
\label{sec-infinite}
As mentioned before, 
the key property of the 
random layer graph is its local weak convergence to
the infinite tree $\bT = (\bV,\bE)$ 
with directed edges,
in which each vertex
has in-degree $2$ and out-degree $2$.  
One vertex of $\bT$ is distinguished as the root.
See Figure 6 later.
The essence of the method is that one can do {\em exact} calculations
within the infinite model which one expects to give the correct
$n \to \infty$ asymptotics for the finite models.
Making the connection rigorous is a challenge we do not address
here, except for brief comments in section \ref{sec-formalize}.
Instead we focus on exhibiting the calculations.

First we copy our finite random network model to

\vspace{0.09in}
\noindent
{\bf The infinite network model.}
Fix a probability distribution on functions $\Phi(v)$
(equivalently: on functions $\phi(v) = v^{-1}\Phi(v)$).
For each $r$ and each edge $e$ of the 
infinite tree $\bT$,
let $\Phi(e,v)$ be chosen independently from this probability distribution.

\vspace{0.09in}
\noindent
A flow $\flow = (f(e))$ in the infinite network is required
only to satisfy the 
``in-flow equals out-flow"
condition at each vertex:
there are no sources and destinations (think of flows from and to
infinitely distant boundaries).
Intuitively, ``normalized volume of flow"
$v(\flow)$
is the average flow per edge over the infinite network.
It is more convenient to interpret this, via the ergodic principle,
as the 
expected value of the flow through a typical edge,
when we require flows to be
{\em invariant}.
Roughly (see \cite{me101} for further discussion)
invariant means that the joint distribution of flow and
edge-capacities and edge-costs is not dependent on the choice 
of root vertex.
In particular, for an invariant flow $\flow$ the quantity
\[ 
v(\flow) = E [f(e)]
\]
does not depend on choice of edge $e$.
This quantity $v(\flow)$ is our definition of 
{\em normalized volume} 
of the flow $\flow$.
Similarly we define the normalized cost associated with a flow
as
\[ c(\flow) =  E [\Phi(e,f(e))]  \]
where again the choice of $e$ does not matter.
Then we study the cost-volume relationship described by the curve 
$c = \psi(v)$:
\[
\psi(v) = \min \{c(\flow): \flow 
\mbox{ an invariant flow with }
v(\flow) = v\}
\]
defined for
\begin{equation}
 0 < v \leq v^* = \max \{v(\flow): \ \flow \mbox{ an invariant flow} \} \leq
 \infty . \label{vstar}
\end{equation}

\subsection{Outline of methodology}
\label{sec-outline}
We are dealing with a minimization-under-constraint problem,
so it is natural to introduce a Lagrange multiplier
$\lambda > 0$
and consider the problem conceptually as
\begin{equation}
\mbox{ 
{\tt minimize }  (cost of flow) - $\lambda \times $(volume of flow)
. } \label{conceptual}
\end{equation}
We analyze this problem on the infinite network
$\bT$ as outlined below.

{\bf Step 1.}
Relative to a reference edge $e^*$, the tree $\bT$
splits into two statistically similar rooted trees
$\bT^+$ and $\bT^-$.

{\bf Step 2.}
On $\bT^+$ consider
\[ X(v) = 
\mbox{
minimum of (\ref{conceptual}) over flows
with $f(e^*) = v$
} \]
measured relative to the $v = 0$ case.

{\bf Step 3.}
$\bT^+$ recursively decomposes into three subtrees,
statistically similar to each other and to $\bT^+$.
The process $X(v)$ is deterministically related to
the corresponding quantities $X_i(v)$ on the subtrees,
and the functions
$\Phi(e_i,v)$
on adjacent edges
$e_i$.
This implies that the distribution of
$(X(v), v \geq 0)$
satisfies a certain
{\em recursive distributional equation}
(RDE), equation 
(\ref{Xtrde}).

{\bf Step 4.}
The flow $f(e^*)$ across $e^*$ in the flow $\flow$ optimizing (\ref{conceptual}) is now determined by
the processes 
$(X^+(v), v \geq 0)$
and
$(X^-(v), v \geq 0)$
on $\bT^+$ and $\bT^-$.

{\bf Step 5.}
From this optimal flow $\flow_\lambda$ we calculate
normalized cost $c(\flow_\lambda)$
and 
normalized volume $v(\flow_\lambda)$
which then determine the cost-volume curve.

\subsection{Discussion of methodology}
\label{sec-formalize}
(a) 
The logic of why we expect this method to give correct answers
is somewhat complicated: here we rephase the outline from \cite{me107} sec. 7.5 
(see also \cite{me101} sec. 5).
Firstly, even though the definition of $X(v)$ is non-rigorous
(the quantity (\ref{conceptual}) equals $\infty - \infty$),
a solution of the RDE 
(\ref{Xtrde}) can be used to process a $\bT$-indexed invariant
random process $(X_e(v))$ with this solution as marginal distribution.
In turn this process 
can be used to define a flow on $\bT$, and the argument which derives
the RDE can (one hopes) be recycled into an argument that this flow is
indeed the optimal flow on the infinite tree.
Identifying this infinite-network optimal flow as the limit
of finite-network optimal flows is the second issue. 
Local weak convergence implies that subsequential weak
limits of optimal finite-$n$ flows are feasible flows on $\bT$, 
but the issue is to show that from the optimal flow on $\bT$
one can synthesize near-optimal flows on the finite networks.
One needs to show that the $\bT$-indexed processes has
a certain ``trivial tail $\sigma$-field' property
(discussed carefully in \cite{me107} under the name {\em endogeny}) .
This implies the optimal flow on an edge is a measurable function of the 
random edge cost-volume functions on other edges.
This enables one to construct quasi-flows (which almost satisfy
the balance requirement at each vertex) on the finite networks,
so then one needs to show that quasi-flows can be converted
to genuine flows with negligible extra cost.

(b) This methodology is fundamentally the same idea as the
non-rigorous
{\em cavity method} 
developed in the 1980s in the study of statistical physics
models of disordered systems. 
See \cite{MParisi03} for the recent survey most useful for our purposes. 
Though intended primarily for study of ``interacting particle" physics models,
it was noted in the 1980s \cite{MPV87} that these methods could be applied also to
combinatorial optimization problems
(matching, traveling salesman) on random points in an artificial
``mean field"
model of geometry 
(complete graph with independent random edge lengths),
and recently have been applied to problems such as random K-SAT \cite{MMZ04}.
Rigorizing cavity method arguments in combinatorial optimization is a project of contemporary interest
in theoretical probability, as yet carried through in only two 
hard problems: see \cite{me94} for the mean-field matching problem  
and \cite{GNS04} for some random graph questions. 
Our example
in section \ref{sec-obstruct}
(maximum density of edge-disjoint
infinite paths in a randomly obstructed 
infinite tree) 
seems a natural next problem for rigorous study.
But in this paper we focus on
demonstrating the range of applicability of the non-rigorous
methodology to network flow problems.
We remark that the third issue in (a) is particular to the  network flow setting, so has not
been studied in previous work.

(c) In most examples we don't expect to be
able to find an explicit analytic solution of the RDE; 
instead we use bootstrap Monte Carlo
(section \ref{sec-bootstrap})
to approximate the solution and derive the numerical results
shown in section \ref{sec-results}.
The theoretical issue of proving uniqueness of solutions is
often difficult.  In the examples in this paper, we
always take $\phi(e,v)$ to be non-decreasing in $v$, so that
$\Phi(e,v)$ is convex in $v$.
By analogy with the deterministic setting
(where a convex function attains its minimum at a unique point)
one might expect convexity to imply uniqueness of solutions of RDEs,
but we do not see any simple general argument.

(d) RDEs are at the center of this formulation
of the cavity method.  As well as their appearance in 
these kind of mean field (disordered network)
optimization problems, they arise in a broad range of
applied probability problems, as illustrated in
the survey \cite{me107}.

\subsection{The general case}
We now show how to implement the section \ref{sec-outline} methodology in
the general infinite network model of section \ref{sec-infinite}.

Fix an edge $e^* = (\vw_-, \vw_+)$ in $\bT$.
(We use $w$ to denote a vertex, since we are using $v$ for {\em volume}).
Delete the other edges at $\vw_-$ and write $\bT^+ = (\bV^+,\bE^+)$
for the component containing $\vw_+$; this is an infinite tree
with the same properties as $\bT$ except that the distinguished vertex
$\vw_-$ has out-degree $1$ and in-degree $0$.
See Figure 6.

\setlength{\unitlength}{0.8in}
\begin{picture}(6,4.4)(0.4,-2)
\put(-0.14,-0.15){$\vw_-$}
\put(0.75,-0.15){$\vw_+$}
\put(0.42,0.09){$e^*$}
\put(0,0){\vector(1,0){0.95}}
\put(1,0){\vector(1,0){0.95}}
\put(1,0){\vector(0,1){0.95}}
\put(1,-1){\vector(0,1){0.95}}
\put(2,0){\vector(1,0){0.45}}
\put(2,0){\vector(0,1){0.45}}
\put(2,-0.5){\vector(0,1){0.45}}
\put(1,1){\vector(0,1){0.45}}
\put(1,1){\vector(1,0){0.45}}
\put(0.5,1){\vector(1,0){0.45}}
\put(0.5,-1){\vector(1,0){0.45}}
\put(1,-1.5){\vector(0,1){0.45}}
\put(1,-1){\vector(1,0){0.45}}
\put(2.5,0){\vector(1,0){0.3}}
\put(2.5,0){\vector(0,1){0.3}}
\put(2.5,-0.33){\vector(0,1){0.3}}
\put(2,-0.5){\vector(1,0){0.3}}
\put(1.66,-0.5){\vector(1,0){0.3}}
\put(2,-0.83){\vector(0,1){0.3}}
\put(1.66,0.5){\vector(1,0){0.3}}
\put(2,0.5){\vector(0,1){0.3}}
\put(2,0.5){\vector(1,0){0.3}}
\put(1.5,1){\vector(1,0){0.3}}
\put(1.5,1){\vector(0,1){0.3}}
\put(1.5,0.67){\vector(0,1){0.3}}
\put(0.66,1.5){\vector(1,0){0.3}}
\put(1,1.5){\vector(0,1){0.3}}
\put(1,1.5){\vector(1,0){0.3}}
\put(0.17,1){\vector(1,0){0.3}}
\put(0.5,1){\vector(0,1){0.3}}
\put(0.5,0.67){\vector(0,1){0.3}}
\put(1.5,-1){\vector(1,0){0.3}}
\put(1.5,-1){\vector(0,1){0.3}}
\put(1.5,-1.33){\vector(0,1){0.3}}
\put(0.66,-1.5){\vector(1,0){0.3}}
\put(1,-1.83){\vector(0,1){0.3}}
\put(1,-1.5){\vector(1,0){0.3}}
\put(0.17,-1){\vector(1,0){0.3}}
\put(0.5,-1){\vector(0,1){0.3}}
\put(0.5,-1.33){\vector(0,1){0.3}}
\put(5.2,0){\vector(1,0){0.75}}
\put(5,0.2){\vector(0,1){0.75}}
\put(5,-1){\vector(0,1){0.75}}
\put(6,0){\vector(1,0){0.45}}
\put(6,0){\vector(0,1){0.45}}
\put(6,-0.5){\vector(0,1){0.45}}
\put(5,1){\vector(0,1){0.45}}
\put(5,1){\vector(1,0){0.45}}
\put(4.5,1){\vector(1,0){0.45}}
\put(4.5,-1){\vector(1,0){0.45}}
\put(5,-1.5){\vector(0,1){0.45}}
\put(5,-1){\vector(1,0){0.45}}
\put(6.5,0){\vector(1,0){0.3}}
\put(6.5,0){\vector(0,1){0.3}}
\put(6.5,-0.33){\vector(0,1){0.3}}
\put(6,-0.5){\vector(1,0){0.3}}
\put(5.66,-0.5){\vector(1,0){0.3}}
\put(6,-0.83){\vector(0,1){0.3}}
\put(5.66,0.5){\vector(1,0){0.3}}
\put(6,0.5){\vector(0,1){0.3}}
\put(6,0.5){\vector(1,0){0.3}}
\put(5.5,1){\vector(1,0){0.3}}
\put(5.5,1){\vector(0,1){0.3}}
\put(5.5,0.67){\vector(0,1){0.3}}
\put(4.66,1.5){\vector(1,0){0.3}}
\put(5,1.5){\vector(0,1){0.3}}
\put(5,1.5){\vector(1,0){0.3}}
\put(4.17,1){\vector(1,0){0.3}}
\put(4.5,1){\vector(0,1){0.3}}
\put(4.5,0.67){\vector(0,1){0.3}}
\put(5.5,-1){\vector(1,0){0.3}}
\put(5.5,-1){\vector(0,1){0.3}}
\put(5.5,-1.33){\vector(0,1){0.3}}
\put(4.66,-1.5){\vector(1,0){0.3}}
\put(5,-1.83){\vector(0,1){0.3}}
\put(5,-1.5){\vector(1,0){0.3}}
\put(4.17,-1){\vector(1,0){0.3}}
\put(4.5,-1){\vector(0,1){0.3}}
\put(4.5,-1.33){\vector(0,1){0.3}}
\put(0.95,2){$\bT^+$}
\put(4.95,2){$\bT_3$}
\put(6.6,0.6){$\bT_2$}
\put(5.7,-1.4){$\bT_1$}
\put(5.49,0.09){$e^*_2$}
\put(5.06,0.44){$e^*_3$}
\put(5.06,-0.56){$e^*_1$}
\put(4.88,-0.05){$\vw_+$}
\end{picture}

\vspace{0.1in}

{\bf Figure 6.}
{\small The tree $\bT^+$ and its recursive decomposition into
$\bT_1,\bT_2,\bT_3$.
}

\vspace{0.13in}
\noindent
Fix a realization of 
edge cost-volume functions
$(\Phi(e,v), e \in \bE^+ \setminus \{e^*\})$.
Let $\FF^+$ be the set of flows $\flow$ on $\bE^+$ which satisfy
the balance constraints at each vertex
except $\vw_-$.
For $0 \leq v < \infty$ define
\begin{eqnarray}
X(v) &=&
\inf_{\flow \in \FF^+: f(e^*)=v}
\sum_{e \in \bE^+, e \neq e^*} (\Phi(e,f(e)) - \lambda f(e)) \nonumber\\
&-&
\inf_{\flow \in \FF^+: f(e^*)=0}
\sum_{e \in \bE^+, e \neq e^*} (\Phi(e,f(e)) - \lambda f(e))  . \label{Xdef}
\end{eqnarray}
As written, one cannot make rigorous sense of these infinite sums.
The heuristic idea is to interpret each sum as a $r \to \infty$
limit of
\[ (\mbox{sum over $e$ within distance $r$ from $e^*$})
\ - \ a_r \]
for normalizing constants $a_r$, and then the constants cancel 
when we subtract to compare the $v>0$ case with the $v=0$ case.
Conceptually, $X(v)$ measures the {\em relative} effect of insisting
that the flow through $e^*$ be exactly $v$.

We next derive the recursion for $X(v)$.
Recall $e^* = (\vw_-,\vw_+)$ is the distinguished edge in $\bT^+$.
Write $e_1^*$ for the other edge directed into $\vw_+$,
and write $e_2^*,e_3^*$ for the two edges directed out of $\vw_+$.
By cutting at $\vw_+$, we can decompose
$\bT^+$ into three subtrees $\bT_1$, $\bT_2$,
$\bT_3$ and the single edge $e^*$, where each $\bT_i$ contains $e_i^*$.
Each $\bT_i$ is isomorphic to $\bT^+$
(with edge-reversal, in the case of $\bT_1$),
and has a distinguished edge $e^*_i$ with $\vw_+$ as the 
exceptional vertex isomorphic to $\vw_-$.
See Figure 6.

On each $\bT_i$ define $X_i(v)$ as at (\ref{Xdef}).
We will show
\begin{eqnarray}
X(v) &=&
\inf_{v+v_1=v_2+v_3} \sum_{i=1}^3 \left(\Phi(e^*_i,v_i) - \lambda v_i + X_i(v_i)\right) 
\nonumber\\
&-&
\inf_{v_1=v_2+v_3} \sum_{i=1}^3 \left(\Phi(e^*_i,v_i) - \lambda v_i + X_i(v_i)\right) 
\label{Xrec}
\end{eqnarray}
To derive this equality, rewrite (\ref{Xdef}) as 
\[ X(v) = \widetilde{X}(v) - \widetilde{X}(0) . \]
In a flow $\flow$ on $\bT^+$ with $f(e^*) = v$, the
flows $f(e_1^*) = v_1, 
f(e_2^*) = v_2,
f(e_3^*) = v_3$
must satisfy $v+v_1 = v_2+v_3$.
For a given value of $v_i$ the contribution to the sum
in (\ref{Xdef}) from edges in $\bT_i$ equals
\[ \Phi(e^*_i,v_i) - \lambda v_i + \widetilde{X}_i(v_i) \]
because we obviously choose the optimal flow on $\bT_i$ for the
given $v_i$. Optimizing over choices of $(v_i)$ gives
\[ \widetilde{X}(v) =
\inf_{v+v_1=v_2+v_3} \sum_{i=1}^3 \left(\Phi(e^*_i,v_i) - \lambda v_i + \widetilde{X}_i(v_i)\right) 
\]
and this leads to (\ref{Xrec}).

A key point is that the subtrees $\bT_i$ with their costs and
capacities
are isomorphic to 
$\bT$ with its costs and
capacities;
and so the three processes $(X_i(v), i = 1,2,3)$
are independent and have the same distribution as $(X(v))$.
Note here that the ``edge-reversal" involved with $e^*_1$ 
makes no difference, since our model is invariant in distribution
under edge-reversal.
Thus (\ref{Xrec}) implies a
{\em recursive distribution equation}
(RDE) for the ``unknown" distribution of 
$X = (X(v), v \geq 0)$, as follows.

\begin{equation}
X \ed F_\lambda (X_1,X_2,X_3,\Phi_1,\Phi_2,\Phi_3) \label{Xtrde}
\end{equation}
where $X_1,X_2,X_3,\Phi_1,\Phi_2,\Phi_3$ are independent;
the $\Phi_i$ are distributed as the edge-cost $\Phi$,
the $X_i$ are distributed as $X$, and
$F_\lambda(x_1(\cdot),x_2(\cdot),x_3(\cdot),\phi_1(\cdot),\phi_2(\cdot),\phi_3(\cdot))$
is the function
\begin{eqnarray*}
 v &\to& 
\inf_{v+v_1=v_2+v_3} \sum_{i=1}^3 \left(\phi(v_i) - \lambda v_i + x_i(v_i)\right) 
\\
&-&
\inf_{v_1=v_2+v_3} \sum_{i=1}^3 \left(\phi(v_i) - \lambda v_i + x_i(v_i)\right) 
 . \end{eqnarray*}
(Here $\phi(\cdot)$ denotes a typical value of $\Phi(\cdot)$.)

Recall the construction of $\bT^+$
as the subtree of $\bT$ on one side of the edge $e^*$.
Construct an opposite subtree $\bT^-$ of $\bT$ by again starting with
the edge $e^* = (\vw_-,\vw_+)$, and now
deleting the other edges at $\vw_+$ to leave $\bT^-$ as the component
containing $\vw_-$.
So $\bT^-$ is isomorphic, under edge-reversal, to $\bT^+$.
Write 
$(X^+(v))$
and
$(X^-(v))$
for the processes (\ref{Xdef}) on $\bT^+$ and $\bT^-$.
Now consider minimizing, over flows $\flow$ on the entire 
tree $\bT$,
the quantity
\[
\sum_{e \in \bE } (\Phi(e,f(e)) - \lambda f(e)) 
 . \]
Any flow $\flow$ decomposes into flows on $\bT^+$ and on $\bT^-$
with the same value of $v = f(e^*)$.
Minimizing the quantity above
for a given value of $v$ gives 
\[ \Phi(e^*,v) - \lambda v + X^+(v) + X^-(v) . \]
Thus the optimal flow is obtained by minimizing over $v$,
and the flow across $e^*$ is
\begin{equation}
f(e^*) = \arg \min_v 
\left( \Phi(e^*,v) - \lambda v + X^+(v) + X^-(v) \right). \label{criterion}
\end{equation}
This optimal flow $\flow = \flow_\lambda$ 
has normalized volume and cost
(section \ref{sec-infinite})
\begin{eqnarray}
v(\flow_\lambda) &=& E [f(e^*) ]
\label{v-lambda}\\
c(\flow_\lambda) 
&=& E [\Phi(e^*,f(e^*))] 
  .
\label{c-lambda}
\end{eqnarray}
This completes the analytic arguments.
We now do bootstrap Monte Carlo (section \ref{sec-bootstrap})
to numerically compute the solution of the RDE 
(\ref{Xtrde}) and then use 
(\ref{criterion},\ref{v-lambda},\ref{c-lambda})
to get the numerical results presented in
sections \ref{sec-traffic}, \ref{sec-capac} and \ref{sec-quadratic}.

\subsection{Unit edge capacities}
\label{sec-UEC}
We now turn to the specialization where each edge has unit capacity
and the cost-per-unit-volume on an edge is constant up to volume $1$:
\begin{eqnarray*}
\phi(e,v) &=& C(e) \quad 0 \leq v \leq 1    \\
&=& \infty \quad v > 1   .
\end{eqnarray*}
So the randomness is supplied only via the i.i.d. edge-costs $C(e)$.
Call a flow $\flow$ with 
\[ f(e) = 0 \mbox{ or $1$ for all } e \]
a $0-1$ flow.
If we consider a random $0-1$ flow ${\bf F} = (F(e))$ then 
the expectations $f(e) = E[F(e)]$ form a flow
with $0 \leq f(e) \leq 1$.
Conversely, any flow 
with $0 \leq f(e) \leq 1$
can be represented as the expectation of a random $0-1$ flow.
It follows that in 
our optimization problem
(\ref{conceptual})
we need only consider $0-1$ flows.
This simplifies the mathematical structure, because
in the RDE (\ref{Xtrde}) we now need consider only 
$X(1)$, which we re-name as $X$.
Looking at (\ref{Xtrde}), there are only three possible
values of $(v_1,v_2,v_3)$ for each case $v = 0,1$:
\[ (v=1):\quad (0,1,0), \ \ (0,0,1), \ \ (1,1,1) \]
\[ (v=0):\quad (0,0,0), \ \ (1,1,0), \ \ (1,0,1) . \]
So (\ref{Xtrde}) becomes a RDE for an unknown distribution 
of a real-valued random variable $X$:
\begin{eqnarray}
X &\ed& \min \left(
X_2+C_2-\lambda, X_3+C_3-\lambda,
\sum_{i=1}^3 (X_i+C_i-\lambda) \right) \nonumber\\
&-& \min \left( 0,
\sum_{i=1,2} (X_i+C_i-\lambda),
\sum_{i=1,3} (X_i+C_i-\lambda)
\right) . \label{Xrde}
\end{eqnarray}
Next, the formula (\ref{criterion}) for optimal flow across $e^*$
says, in the present setting, that the optimal flow has unit
flow across $e^*$ iff the arg min in (\ref{criterion}) equals
$1$ instead of $0$, that is iff
\[ 0 > C(e^*) - \lambda + X^+ + X_- \]
where $X^+$ and $X^-$ are the independent copies of $X$
associated with $\bT^+$ and $\bT^-$.
Thus we get the 
{\em inclusion criterion}:
$e^*$ is in the optimal flow iff
\begin{equation}
C(e^*) < \lambda - X^+ - X^- .
\label{criterion-2}
\end{equation}
So the normalized volume and cost of the
optimal flow $\flow_\lambda$ are
\begin{eqnarray}
v(\flow_\lambda) &=& 
P(C(e^*) < \lambda - X^+ - X^- ) 
\label{v-lambda-2}\\
c(\flow_\lambda) &=& E \left[ C(e^*)1_{(
C(e^*) < \lambda - X^+ - X^- ) }
\right] 
. \label{c-lambda-2}
\end{eqnarray}
As before, we can now use bootstrap Monte Carlo to
solve 
(\ref{Xrde}) numerically, and then use
(\ref{criterion-2},\ref{v-lambda-2},\ref{c-lambda-2})
to compute the curve in section \ref{sec-scaling}.

\subsection{Randomly obstructed networks}
To analyze the section \ref{sec-obstruct} model, recall
(\ref{Cflow}) that we can interpret it as the case of unit capacity edges
with costs $C$ such that
\[ P(C = 1) = p; \quad P(C = \infty) = 1-p .  \] 
Looking back at (\ref{conceptual}) we see that, because
edges-costs are either $1$ or $\infty$, one must get the
maximum volume flow for an arbitrary choice of $\lambda > 1$,
and the solution $X$ of (\ref{Xrde}) should be supported on
multiples of $1 - \lambda$.
Examining (\ref{Xrde}), we see 
the latter is correct.
Setting $X = Z(\lambda - 1)$ in (\ref{Xrde}) leads to the RDE
(not depending on $\lambda$)
\begin{eqnarray}
Z &\ed& \max \left(
Z_2+B_2, Z_3+B_3,
\sum_{i=1}^3 (Z_i+B_i) \right) \nonumber\\
&-& \max \left( 0,
\sum_{i=1,2} (Z_i+B_i),
\sum_{i=1,3} (Z_i+B_i)
\right) \label{Zrde}
\end{eqnarray}
where $Z$ has unknown distribution on 
$\{- \infty\} \cup \Zbold$ 
and where $(B_i)$ are independent with 
$P(B = 1) = p, \ P(B = -\infty) = 1-p$.
In terms of two copies $Z^+, Z^-$ of the solution of this RDE, 
(\ref{v-lambda-2}) implies the formula
\begin{equation}
v^*(p) = P(Z^+ + Z^- > -1) .
\end{equation}
As usual, we solve this numerically by bootstrap Monte Carlo 
to obtain the curve shown in Figure 5.

\subsection{Bootstrap Monte Carlo}
\label{sec-bootstrap}
The abstract structure of a RDE is
\[X \ed g(\xi,X_i, \ i \geq 1) \]
where $g(\cdot)$ and the distribution of $\xi$ are given,
and where $(X_i, i \geq 1)$ are independent copies of an
``unknown" distribution $X$.
Here $X$ and $\xi$ can take values in arbitrary spaces.
Equivalently, an RDE is a fixed-point equation for a 
map $\mu \to T(\mu)$ on probability distributions, where
\[ T(\mbox{dist}(X)) = \mbox{dist}(
g(\xi,X_i, \ i \geq 1) ). \]
The {\em bootstrap Monte Carlo} method provides a very easy to
implement and essentially problem-independent method
to seek solutions.
Start with a list of $N$ numbers 
with some empirical distribution $\mu_0$.
Regard these as ``generation $0$" individuals
$(X^0_i, 1 \leq i \leq N)$.
Then $T(\mu_0)$ can be approximated as the empirical distribution
$\mu_1$ of $N$ ``generation $1$" individuals $(X^1_i, \ 1 \leq i \leq N)$, each obtained
independently via the following procedure.
Take $\xi$ with the prescribed distribution,
take $I_1,I_2,\ldots$ independent uniform on $\{1,2,\ldots,N\}$
and set
\begin{equation}
 X^1_i = g(\xi, X^0_{I_1},X^0_{I_2},\ldots ) . \label{MC-recurse}
\end{equation}
Repeating for some number $G$ of generations 
lets one see
whether $T^n(\mu_0)$ settles down to a solution of the RDE.

Experience with a range of RDEs indicates that taking
$N = 200,000$ as ``population size" and iterating through
$G = 200$ ``generations" gives reliable solutions.
When dist($X$) is just a distribution on the real line,
this procedure requires only $4 \times 10^7$ evaluations of
the form (\ref{MC-recurse}), which is computationally easy
when $g(\cdot)$ is simple to evaluate.
This is the situation for Table 1 and Figure 5.
But in the general setting of this paper, where the unknown
distribution is of a {\em process} $X = X(v)$, 
and the function $g$ involves minimizing over
choices $(v_1,v_2,v_3)$ as at (\ref{Xtrde}), 
the computational problem becomes harder.
Our results in Figures 2,3,5 used a crude implementation
where we represented $X$ via evaluation at 60 grid points $(X(u_1), X(u_2),\ldots, X(u_{60})$.
So one evaluation of (\ref{MC-recurse}) requires 
$60^3$ steps, meaning that using the previous values 
of $N$ and $G$ would require more than $8 \times 10^{12}$ steps.
This being infeasible, we used smaller values of $N$ and $G$,
and the resulting ``sampling error" is visible in the irregularities
in Figures 2,3,5,
where we plotted actual data rather than a smoothed curve.

\section{Other analysis}
\subsection{The linear case and first passage percolation}
\label{sec-linear}
The {\em linear case} is the case
\[ 
\Phi(e,v) = \kappa(e) \ v, \quad 0 \leq v < \infty \]
where the $\kappa(e)$ are i.i.d. as $e$ varies.
In other words, the cost-per-unit-flow
$\phi(e,v) = \kappa(e)$
does not depend on volume of flow.
This has the obvious scaling property that if
$\flow$ has 
$v(\flow) = v_0, \ c(\flow) = c_0$
then a scaled flow $\alpha \flow$ has
$v(\alpha \flow) = \alpha v_0, \ c(\alpha \flow) = \alpha c_0$.
So the network cost-per-unit-volume function must be of the form
\[
c= \psi(v) = \bar{\kappa} 
\]
where $\bar{\kappa}$
depends on the distribution of $\kappa(e)$.
It is easy to see that
$\bar{\kappa}$ can be identified with the
time constant for 
first-passage percolation on $\bT$, that is the limit
\[ n^{-1}  \min_{ {\rm path\ } \root = w_0,w_1,\ldots,w_n}
\sum_{i=1}^n \kappa(w_{i-1},w_i) 
\to \bar{\kappa} \mbox{ a.s. as } n \to \infty . \]
It is well known that,
as a specialization of general results for branching
random walk
(cf. \cite{dur91} Example 6.7.3),
$\bar{\kappa}$ can be calculated as the solution of
\begin{equation}
\inf_{\theta > 0} \left(
\log E [\exp(-\theta \kappa(e))] \ + \theta \bar{\kappa} \right)
= \log 1/2 . \label{eq-brw}
\end{equation}
However, this linear case is (from our viewpoint) 
degenerate in the sense that there is no flow on $\bT$ attaining
the infimum of normalized cost for given normalized volume.
Instead, there is a sequence of flows which assign zero volume to most
edges and assign larger and larger volumes to paths whose average
edge-cost is closer and closer to $\bar{\kappa}$.
Our non-linear examples, and our methodology for analyzing them, 
rest upon the idea that optimal flows on $\bT$ are actually attained by
some minimizing flow $\flow$.

On the other hand, the linear case does tell us something about the low-volume
limit of the general case.
Suppose
\begin{equation}
v \to \phi(e,v) \mbox{ is increasing}; \quad 
\kappa(e):= \phi(e,0+) > 0 .
\label{phi-lowflow}
\end{equation}
Then the low-volume limit of the network cost-per-unit-flow
function will be
\begin{equation}
\psi(0+) = 
\mbox{ the solution $\bar{\kappa}$ of (\ref{eq-brw}).}
\label{eq-psi0+}
\end{equation}
To see why, fix $\eps > 0$ and consider a flow in the linear case
with normalized volume $1$ and normalized cost $\bar{\kappa} + \eps$.
The scaled flow with normalized volume $v$ has, 
in the linear case, normalized cost 
$\psi(v) = (\bar{\kappa} + \eps)v$.
So the same flow in the general case has normalized cost
$\psi(v) \sim (\bar{\kappa} + \eps)v$
as $v \downarrow 0$, by (\ref{phi-lowflow}).

\subsection{The \CD scheme}
\label{sec-greedy}
As mentioned in section \ref{sec-quadratic},
one can consider the 
decentralized routing scheme in which customers leaving each vertex
choose the cheaper out-edge to traverse next.
Writing $e_1,e_2$ for the out-edges at a vertex,
and $\phi(e_i,v_i)$ for the cost-per-unit-volume
functions on those edges, 
the effect of this {\em \CD} scheme is to adjust the flows $v_i$
so that these marginal costs are equal.
That is, if the total in-flow equals $v$, then the out-flows
$v_1, v_2$ are determined by
\begin{equation}
\phi(e_1,v_1) = \phi(e_2,v_2); \quad 
v_1 + v_2 = v .
\end{equation}
We can study the resulting flow in our infinite-network model,
though as noted earlier it is not so clear how results pull back
to the finite network model.
Given $v$ and $\phi_1(\cdot), \phi_2(\cdot)$,
consider the solution $(v_1,v_2)$ of the analog of the equation above:
\begin{equation}
\phi_1(v_1) = \phi_2(v_2); \quad 
v_1 + v_2 = v .
\label{phitt}
\end{equation}
Write
\begin{eqnarray*}
T(v,\phi_1,\phi_2) &=& v_1 \\
W(v,\phi_1,\phi_2) &=& \phi_1(v_1) .
\end{eqnarray*}
It is clear that the flow $Y$ across a typical edge
will satisfy the RDE
\begin{equation}
Y \ed T(Y_1+Y_2,\phi_1,\phi_2)
\label{RDE-greedy}
\end{equation}
where $\phi_i(\cdot)$ denote independent choices from the random
cost-per-unit-volume function
$\phi(v) = \Phi(v)/v$
in the model description.
To see (\ref{RDE-greedy}), note that the flow into a typical vertex
is distributed as $Y_1 + Y_2$, so that 
$ T(Y_1+Y_2,\phi_1,\phi_2)$
represents the flow along one out-edge.

Analogous to the curve
$c = \bar{\Psi}(v)$
giving the normalized cost-volume relationship for
{\em optimal} flow through the infinite network,
there is a curve
$c = \bar{G}(v)$
giving the normalized cost-volume relationship for
the \CD scheme.
We expect (\ref{RDE-greedy}) to have a one-parameter family of solutions
corresponding to different values of $E[Y]$.
In terms of these solution, the curve is
\begin{eqnarray}
v &=&E[Y] = E[
 T(Y_1+Y_2,\phi_1,\phi_2)]
\label{ge1}\\ c = \bar{G}(v) &=&
E \left[ T(Y_1+Y_2,\phi_1,\phi_2)
 W(Y_1+Y_2,\phi_1,\phi_2)
\right]
\label{ge2}
\end{eqnarray}
because the flow of volume
$Y = T(Y_1+Y_2,\phi_1,\phi_2)$
along the typical out-edge has
cost-per-unit-volume equal to
$ W(Y_1+Y_2,\phi_1,\phi_2)$.

We now specialize to the ``quadratic cost" case of section \ref{sec-quadratic}.
Here
\[ \phi_i(v) = \kappa_i v \]
and so we can solve (\ref{phitt}) to get
\begin{eqnarray*}
T(v,\phi_1,\phi_2) &=& \frac{\kappa_2 v}{\kappa_1 + \kappa_2}\\
W(v,\phi_1,\phi_2) &=& \frac{\kappa_1 \kappa_2 v}{\kappa_1 + \kappa_2}.
\end{eqnarray*}
The RDE (\ref{RDE-greedy}) becomes
\begin{equation}
Y \ed \frac{\kappa_2}{\kappa_1 + \kappa_2} \ (Y_1+Y_2) .
\label{RDE-g}
\end{equation}
Specializing (\ref{ge1},\ref{ge2}) we see that the network
cost-volume curve will be
\[ c = \bar{G}(v) = \bar{g}v^2 \]
\begin{equation}
\bar{g} = E \left[
\frac{\kappa_1 \kappa_2^2}{(\kappa_1+\kappa_2)^2} \ 
(Y_1+Y_2)^2 \right]
\label{barg}
\end{equation}
where $Y$ is the solution of (\ref{RDE-g}) with $E[Y] = 1$.
Note the random variables in (\ref{barg}) are all independent.

We now consider the special case where the distribution
of $\kappa$ is $Gamma(a,a)$ for some $0<a<\infty$. 
We will show (as stated in section \ref{sec-quadratic})
that in this case $\bar{g} = 1$.
Recall the Gamma($a,a$) distribution has mean $1$ and variance $1/a$.
It is a classical fact that
\begin{equation}
 \frac{\kappa_2}{\kappa_1 + \kappa_2}
\mbox{ and } \kappa_1 + \kappa_2 
\mbox{ are independent.} \label{KGind}
\end{equation}
It follows that the solution $Y$ of (\ref{RDE-g}) is the same
Gamma$(a,a)$ distribution, because for such $Y$
\[ \frac{\kappa_2}{\kappa_1 + \kappa_2}
(Y_1+Y_2) \ed 
 \frac{\kappa_2}{\kappa_1 + \kappa_2}
(\kappa_1 + \kappa_2) 
= \kappa_2 \ed Y . \]
It remains to evaluate $\bar{g}$.
First observe
\[ E[(Y_1+Y_2)^2] = 4 + 2 \ \var (Y)
= 4 + 2a^{-1} . \]
Next, writing
$\beta = 
 \frac{\kappa_2}{\kappa_1 + \kappa_2}
$ and $\Lambda = \kappa_1 + \kappa_2$
for the independent random variables in (\ref{KGind}),
we can write
\[ 
\frac{\kappa_1 \kappa_2^2}{(\kappa_1+\kappa_2)^2} \ 
= \kappa_1 \beta^2
= \Lambda (1-\beta) \beta^2 . \]
Since $E[\Lambda] = 2$ we 
can insert into (\ref{barg}) to get
\[ \bar{g} = (8 + 4a^{-1}) E[\beta^2 - \beta^3] . \]
But $\beta$ has the Beta($a,a$) density
\[ f(x) = x^{a-1}(1-x)^{a-1}\Gamma(2a)/\Gamma^2(a), \quad 0<x<1 \]
where $\Gamma(\cdot)$ is the gamma function.
The $k$'th moment of this distribution equals
$\frac{\Gamma(2a)\Gamma(a+k)}{\Gamma(2a_k)\Gamma(a)}$,
and a brief calculation shows
\[ E[\beta^2 - \beta^3] = \frac{a}{4(2a+1)} \]
so that indeed $\bar{g} = 1$.

\subsection{Endpoint behavior in the randomly obstructed network model}
\label{sec-ROends}
The endpoint behavior observed numerically in Figure 5
is not too hard to understand theoretically in the infinite tree model, as we now outline.
First we assert
\[ v^*(p) \leq p \theta^2(p) \]
where $\theta(p)$ is the non-extinction probability for the Galton-Watson
branching process with Binomial($2,p$) offspring.
Recalling that we need only consider $0-1$ flows, this is clear
because in order to have a unit flow through $e$, 
we need $e$ itself to be non-obstructed (probability $p$)
and we need there to exist infinite non-obstructed paths
starting from each end-vertex of $e$ 
(probability $\theta(p)$ each).
An elementary calculation gives
$\theta(p) \sim 8(p-1/2)$
as $p \downarrow 1/2$,
and so we deduce
\[ v^*(p) \leq (32 + o(1))(p-\sfrac{1}{2})^2 \mbox{ as } 
p \downarrow \sfrac{1}{2} . \]
Turning to the case where $p$ is close to $1$,
dual to the optimal flow is the complementary ``$0$-flow" consisting
of the set of edges with no flow; this set must make an edge-disjoint
collection of doubly infinite paths containing every
obstructed edge.
By considering where the obstructed edges appear in this $0$-flow
it is easy to see the identity
\begin{quote}
$\frac{1-v^*(p)}{1-p} = $ 
mean number of edges traversed in the optimal $0$-flow, 
starting at a typical obstructed edge, until the next obstructed 
edge is reached.
\end{quote}
Now this mean is $\geq E[M]$ where $M$ is the mean number of
edges, starting at the root of $\bT$ and following directed edges,
needed to reach the {\em closest} obstructed edge.
Because there are
$2+ 2^2 + \ldots + 2^{m-1}$
edges at distance $<m$ we see
\[ E[M]= \sum_{m=1}^\infty P(M \geq m) 
= \sum_{m=1}^\infty p^
{2+ 2^2 + \ldots + 2^{m-1}}
= \log_2 \sfrac{1}{1-p} \pm 0(1) 
\mbox{ as } p \uparrow 1. \]
This shows
\[ 1 - v^*(p) \geq (\log_2 \sfrac{1}{1-p} \ - O(1)) (1-p)
\mbox{ as } p \uparrow 1. \]

These arguments give the easier directions of inequalities, but
proving complementary bounds
\[ v^*(p) \geq a_0(p-\sfrac{1}{2})^2 \mbox{ as } 
p \downarrow \sfrac{1}{2}  
\quad \quad (\mbox{ for some } a_0 > 0)
\]
\[ 1 - v^*(p) \leq a_1 (\log_2 \sfrac{1}{1-p} ) (1-p)
\mbox{ as } p \uparrow 1 
\quad \quad (\mbox{ for some } a_1 < \infty)
\]
is surely within the scope of known methods of theoretical probabilistic
combinatorics, though we have not tried to write down details.

\section{Discussion}
\label{sec-discuss}
\subsection{Other underlying graph models}
The calculations go over with only straightforward 
changes to any model which is ``locally tree-like" in the sense of local weak convergence to 
{\em some} 
limit infinite (maybe random) tree.
Such models include\\
(i) the classical Erd\H{o}s-R\'{e}nyi random graph model \cite{bol85,jansonlr00} (more precisely, the giant component in the sparse supercritical regime);\\
(ii) recent ``complex network" models designed to have power-law degree
distributions \cite{newman-survey}.\\
On the other hand, models which pay attention to Euclidean
geometry of vertex positions ,
such as {\em random geometric graphs} \cite{penrose03},
are not locally tree-like and rarely permit analytical derivation
of exact limit formulas.

\subsection{Other flow and routing problems}
From the algorithmic viewpoint, finding optimal routes through
a realization of
the random layer network with cost-volume functions on each edge 
is not easy.  
It is therefore remarkable that one can give a theoretical analysis of 
costs of the optimal routing without any consideration 
whatsoever of algorithmic issues!  
Of course, our focus on the global optima is unrealistic, and it would
be interesting to use our models as a testbed for comparative analysis of different
distributed routing algorithms.

\paragraph{Acknowledgement.}
We thank Frank Kelly for helpful discussions.

\newpage

\def\cprime{$'$} \def\polhk#1{\setbox0=\hbox{#1}{\ooalign{\hidewidth
  \lower1.5ex\hbox{`}\hidewidth\crcr\unhbox0}}} \def\cprime{$'$}
  \def\cprime{$'$} \def\cprime{$'$}
  \def\polhk#1{\setbox0=\hbox{#1}{\ooalign{\hidewidth
  \lower1.5ex\hbox{`}\hidewidth\crcr\unhbox0}}} \def\cprime{$'$}
  \def\cprime{$'$} \def\polhk#1{\setbox0=\hbox{#1}{\ooalign{\hidewidth
  \lower1.5ex\hbox{`}\hidewidth\crcr\unhbox0}}} \def\cprime{$'$}
  \def\cprime{$'$} \def\cydot{\leavevmode\raise.4ex\hbox{.}} \def\cprime{$'$}
  \def\cprime{$'$} \def\cprime{$'$} \def\cprime{$'$} \def\cprime{$'$}
  \def\cprime{$'$} \def\cprime{$'$} \def\cprime{$'$} \def\cprime{$'$}
  \def\cprime{$'$}

\end{document}